\documentclass[prl,twocolumn,showpacs,amssymb]{revtex4}

\usepackage{psfrag,graphicx}
\usepackage{dcolumn}
\usepackage{bm}
\usepackage{amsfonts,amssymb,amsmath}        

\usepackage{xcolor}

\newcommand{\be}{\begin{equation}}
\newcommand{\ee}{\end{equation}}
\newcommand{\bq}{\begin{eqnarray}}
\newcommand{\eq}{\end{eqnarray}}

\newcommand{\rf}[1]{(\ref{#1})}

\newcommand{\GAMMA}{\mbox{\boldmath${\gamma}$}}

\newcommand{\p}{{\bf p}}

\begin{document}

\title{Non-Abelian Chern-Simons Theory from a Hubbard-like Model} 

\author{Giandomenico Palumbo}
\affiliation{School of Physics and Astronomy, University of Leeds, Leeds, LS2 9JT, United Kingdom}
\author{Jiannis K. Pachos}
\affiliation{School of Physics and Astronomy, University of Leeds, Leeds, LS2 9JT, United Kingdom}

\date{\today}


\begin{abstract}

Here, we provide a simple Hubbard-like model of spin-$1/2$ fermions that gives rise to the SU(2) symmetric Thirring model that is equivalent, in the low-energy limit, to Yang-Mills-Chern-Simons model. First, we identify the regime that simulates the SU(2) Yang-Mills theory. Then, we suitably extend this model so that it gives rise to the SU(2) level $k$ Chern-Simons theory with $k\geq2$ that can support non-Abelian anyons. This is achieved by introducing multiple fermionic species and modifying the Thirring interactions, while preserving the SU(2) symmetry. Our proposal provides the means to theoretically and experimentally probe non-Abelian SU(2) level $k$ topological phases.

\end{abstract}

\maketitle

Interacting systems are in general too hard to track analytically. An interesting approach is to employ low-dimensional interacting relativistic quantum field theories at zero temperature for which bosonisation can be applied. Some of these theories can be simultaneously analytically tractable {\em and} amendable to experimental verification, e.g. with cold atoms. The $(2+1)$-dimensional Thirring model \cite{Gomes}, that describes interacting Dirac fermions, provides such example. If the interaction term possesses U(1) symmetry then the model is equivalent through bosonisation to the Maxwell-Chern-Simons theory \cite{Fradkin}. If the interaction term is SU(2) symmetric then the model can be described by Yang-Mills-Chern-Simons theory \cite{Fradkin2}. Unfortunately, the anyons supported by this model are Abelian, namely SU(2) level $k=1$ anyons. 

The goal of this report is twofold. First, to present a Hubbard-like model of spin-$1/2$ fermions that gives rise in the continuum limit to the SU(2) symmetric Thirring model. In particular, we identify the coupling regime where the Yang-Mills theory is predominant in the bosonised version of the model. Hence, the model could serve as a quantum simulator for demonstrating confinement in $2+1$ dimensions, e.g. with current cold atom technology. 
Although quantum simulators for lattice Yang-Mills theory in cold 
atomic systems have been recently proposed in \cite{Lewenstein, Zoller, Cirac}, our model simulates a continuum non-Abelian gauge theory. 
Second, we employ multiple species of fermions so that the low energy of the model is described by the SU(2) level $k\geq2$ Chern-Simons theory. This theory can support non-Abelian anyons such as Ising or Fibonacci anyons. Hence, its physical realisation can serve for the implementation of topological quantum computation \cite{Pachos12}. 

A finite temperature implementation of our work is also possible. Indeed, the non-Abelian Chern-Simons theory can be induced by fermions also at finite temperature \cite{Shapo2, Shapo3}. This analysis goes beyond the scope of the paper and it will be left to future work.

\begin{figure}
\includegraphics[width=0.47\textwidth]{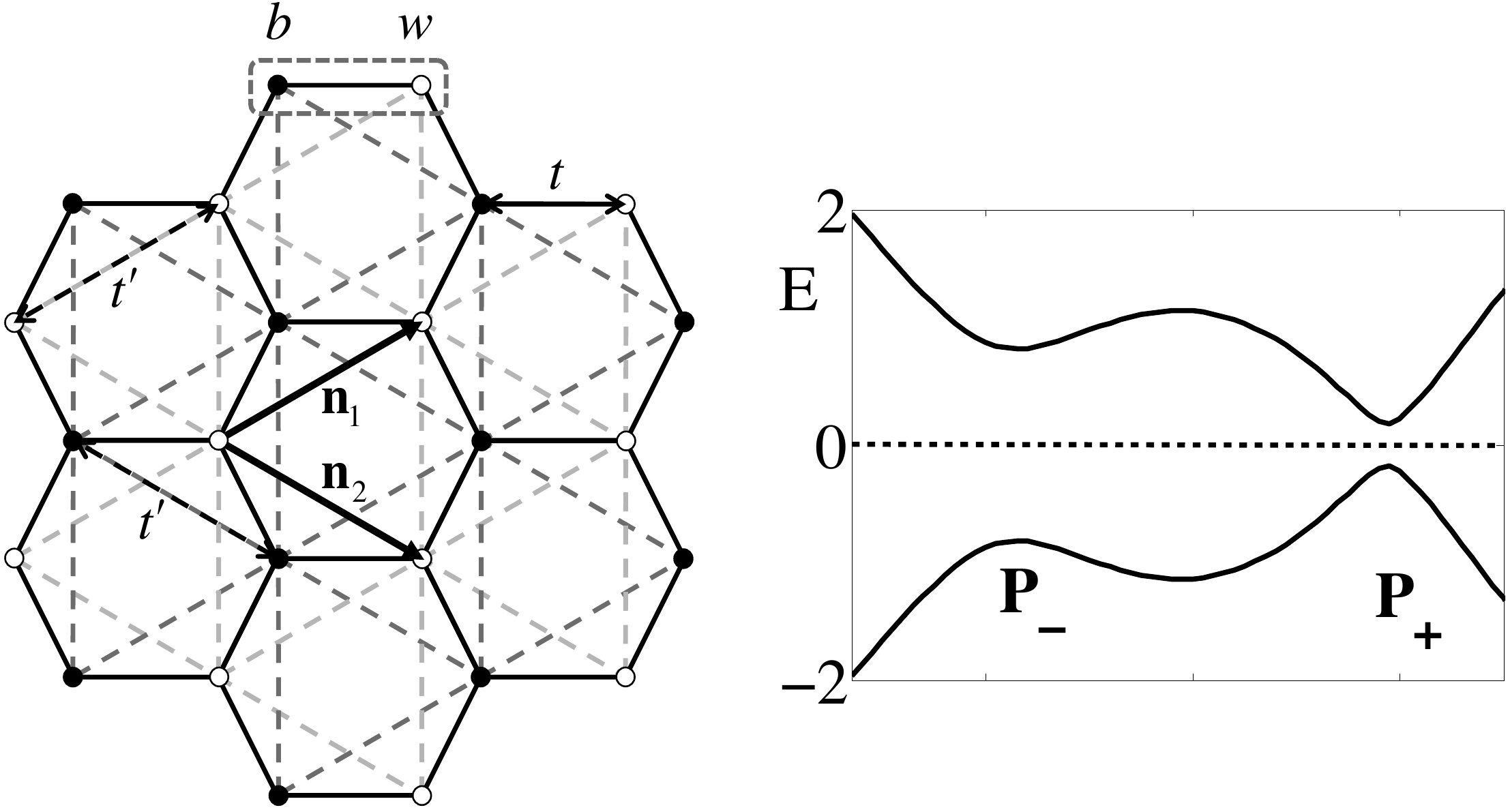}
\caption{\label{fig:lattice} (Left) The honeycomb lattice with its unit cell consisting of two sites, $b$ and $w$. Fermions at a certain site tunnel to their neighbouring and next-to-neighbouring sites, with coupling $t$ and $t'$, respectively. The phase factor $\chi_{{\bf i},{\bf j}} = \pm i$ in Hamiltonian (\ref{Ham1}) has $+$ sign when the link $\langle\langle i,j\rangle\rangle$ points along the directions of ${\bf n}_1$, ${\bf n}_2$ or ${\bf n}_1-{\bf n}_2$ and $-$ sign otherwise, where ${\bf n}_1= (3/2,\sqrt{3}/2)$ and ${\bf n}_2= (3/2,-\sqrt{3}/2)$. (Right) The energy dispersion relation along $p_y$ where both Fermi points, ${\bf P}_\pm$, reside for generic values of the couplings. The corresponding energy gaps, $\Delta E_\pm$, can be independently tuned. }
\end{figure}

Our starting point is a tight-binding model with low energy behaviour described by the SU(2) symmetric Thirring model in $2+1$ dimensions. The Thirring model comprises of interacting relativistic Dirac fermions. To simulate it we employ tight-binding fermions in a honeycomb lattice configuration, as shown in Fig. \ref{fig:lattice} (Left). We introduce the Hubbard-like Hamiltonian
\bq
H = \!\!&-&\!\! t \sum_{\langle {\bf i},{\bf j}\rangle, s} (b^\dagger_{s,{\bf i}} w^{}_{s,{\bf j}} +w^\dagger_{s,{\bf i}}b^{}_{s,{\bf j}}) -\mu \sum_{{\bf i}, s} (n^b_{s,{\bf i}} - n^w_{s,{\bf i}}) \nonumber\\
\!\!&-&\!\! \sum_{\langle\langle{\bf i},{\bf j}\rangle\rangle, s} \chi_{{\bf i},{\bf j}} \,t' ( b^\dagger_{s,{\bf i}}b^{}_{s,{\bf j}} - w^\dagger_{s,{\bf i}}w^{}_{s ,{\bf j}})\nonumber \\
\!\!&+&\!\!U\Big(\sum_{{\bf i}, s,s'}n^b_{s,{\bf i}} n^w_{s',{\bf i}} -\sum_{\bf i, \alpha}n^\alpha_{\uparrow,{\bf i}} n^\alpha_{\downarrow,{\bf i}} \Big),
\nonumber \\
\label{Ham1}
\eq
where $n^\alpha_s = \alpha^\dagger_s\alpha_s$ is the population of particle $\alpha=b,w$, distinguished by their position in the unit cell, with spin $s=\uparrow,\downarrow$. The phase factor $\chi_{{\bf i},{\bf j}} = \pm i$ is defined in Fig.~\ref{fig:lattice} (Left). The $t$-term of the Hamiltonian corresponds to tunnelling along the honeycomb lattice. In the continuum limit it gives rise to two massless Dirac fermions corresponding to the Fermi points ${\bf P}_\pm = (0,\pm 4\pi /(3\sqrt{3}))$ in Cartesian coordinates. The chemical potential $\mu$-term and the next-to-nearest tunnelling $t'$-term give rise to energy gaps at the two Fermi points of the form
\be
\Delta E_\pm = 2|-\mu\pm \sqrt{3}t'|.
\ee
For $\Delta E_+\ll \Delta E_-$, as shown in Fig. \ref{fig:lattice} (Right), we can adiabatically eliminate the ${\bf P}_-$ Fermi point from the low energy dynamics of the system \cite{Giandomenico}. Hence, we  can isolate the dynamics of the single Fermi point ${\bf P}_+$. An alternatively approach to the adiabatic elimination is to consider three-dimensional topological insulator with an isolated Dirac cone  at its boundary \cite{Ryu}. Introducing suitable boundary fields generates an energy gap, so the surface state can be effectively described by a massive Dirac fermion.

\begin{figure}
\includegraphics[width=0.47\textwidth]{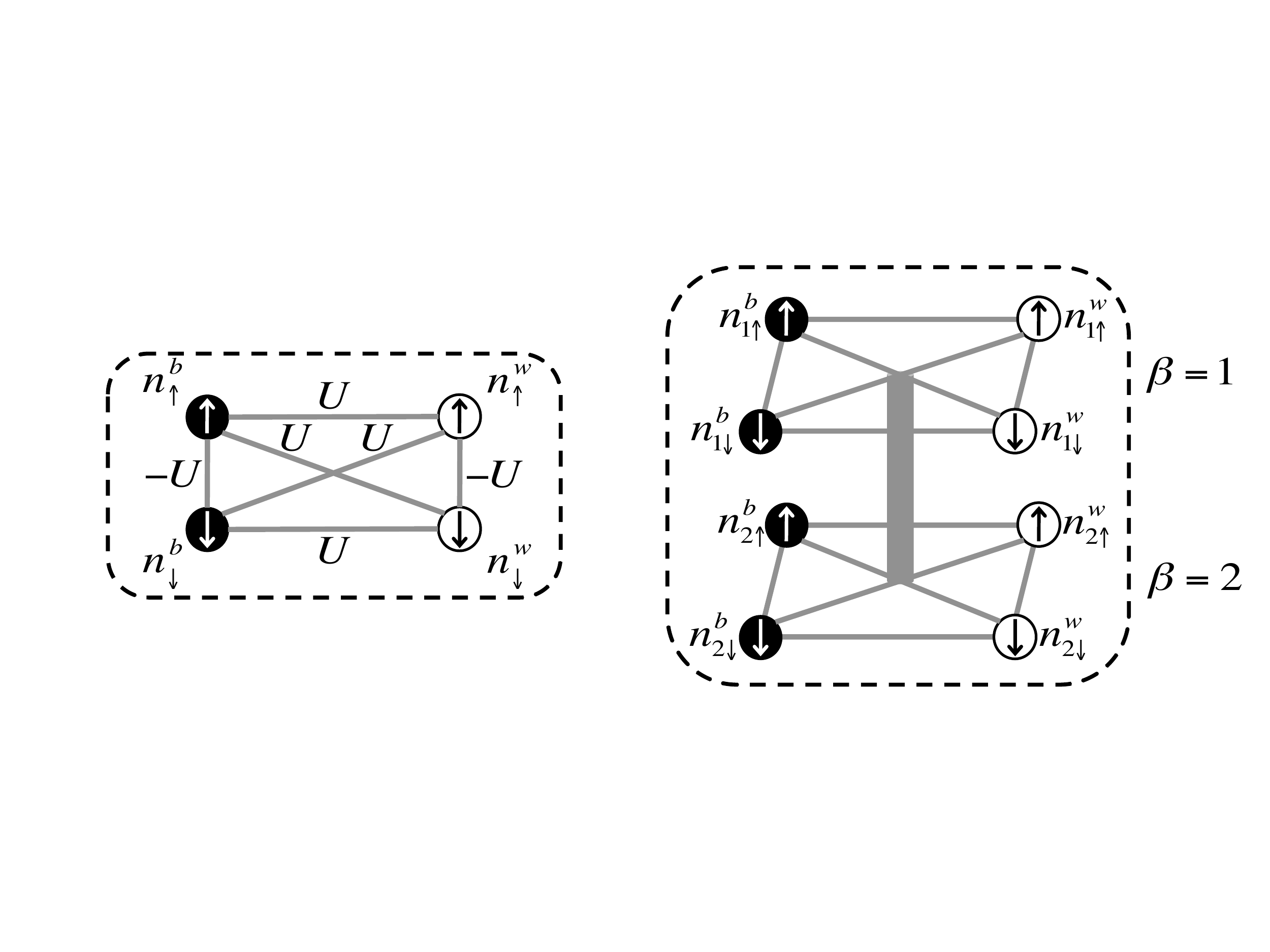}
\caption{\label{fig:interactions} The fermionic interactions, given by grey lines, within a single unit cell that includes one black and one white site (see Fig. \ref{fig:lattice}). (Left) The interactions for the single fermionic species model between populations $n^\alpha_s = \alpha^\dagger_s\alpha_s$ with $\alpha=b,w$, $s=\uparrow,\downarrow$ and strength $\pm U$. (Right) The interactions for the two fermionic species model. We can consider this as a bilayered system with the interactions between the populations $n^\alpha_{\beta s} = \alpha^\dagger_{\beta s}\alpha_{\beta s}$ with $\alpha=b,w$, $s=\uparrow,\downarrow$ and $\beta=1,2$ given explicitly by Hamiltonian (\ref{eqn:extended}).}
\end{figure}

By introducing the spinor $\psi = (\psi_\uparrow,\psi_\downarrow)^T = (b_\uparrow,w_\uparrow,b_\downarrow,w_\downarrow)^T$ with $\psi_s = (b_s,w_s)^T$ and $s=\uparrow,\downarrow$, we can write the interaction $U$-term of Hamiltonian \ref{Ham1} in the form ${2 \over 3} U (\overline \psi T^a\gamma^\mu \psi) (\overline \psi T^a\gamma_\mu \psi)$ that acts locally within the unit cell. Here $\overline \psi = \psi^\dagger\gamma_z$, $\gamma_\mu = \sigma_\mu\otimes \mathbb{I}_{2} $ for $\mu = x,y,z$ are $4\times4$ Euclidean Dirac matrices written in terms of the Pauli matrices, where $\mathbb{I}_2$ acts on the spin subspace, and $T^a = \sigma^a/2$, for $a=x,y,z$, are the generators of SU(2). The arrangement of the tight-binding interactions that give rise to the self-interaction of the Dirac fermion is shown in Fig. \ref{fig:interactions} (Left).

In the low energy limit the behaviour of the model around ${\bf P}_+$ is given by the Hamiltonian
\be
H = \int d^3 x \Big[\psi^\dagger (v \gamma_z\GAMMA\cdot \p+\gamma_z M v^2)\psi + {g^2 \over 2} j^{a\mu} j^a_{\mu} \Big],
\label{eqn:ThHam}
\ee
where $j^{a\mu} =\overline \psi T^a\gamma^\mu \psi$, $v={3\over 2}t$, $Mv^2 =-\mu+ \sqrt{3}t'$, $g^{2}={4\over 3}{U}$. For simplicity we take from now on $v=1$. Hamiltonian \rf{eqn:ThHam} corresponds to the $(2+1)$-dimensional Thirring model with SU(2) symmetry. This non-Abelian symmetry is manifested by the invariance of the Hamiltonian under transformations of the spinor $\psi^V_s = V_{ss'}\psi_{s'}$, for $V\in$ SU(2). Note that this symmetry of the interacting term is also exact in the discrete model.

It is known that in $2+1$ or higher dimensions even the Abelian Thirring model  is perturbatively non-renormalisable. Nevertheless, it has been shown to become renormalisable in the non-perturbative large-$N$ limit~\cite{Parisi,Gies}. In our case we are only interested in the low energy sector of the tight-binding model and, consequently, in the infrared limit of the corresponding SU(2) Thirring model. In the following we show how this model maps to a renormalisable gauge theory to leading order in $1/M$  \cite{Fradkin2}. This mass fixes the validity energy range of our effective theory. 

We show now the connection between the SU(2) symmetric Thirring model and the Yang-Mills-Chern-Simons theory \cite{Fradkin2}. To proceed we employ the path integral formalism with Euclidean signature. The Non-Abelian Thirring action that corresponds to Hamiltonian (\ref{eqn:ThHam}) is given by
\begin{eqnarray}
S_\mathrm{Th}=\int d^{3}x \;
\left[\overline{\psi}(\displaystyle{\not}\partial-M)\psi-\frac{g^{2}}{2}
j^{a\mu}j^a_{\mu}\right],
\end{eqnarray}
and the corresponding partition function is defined as $Z_\mathrm{Th}=\int\mathcal{D}\overline{\psi}\mathcal{D}\psi\; e^{- S_\mathrm{Th}}$. To treat the interaction term that is quartic in the fermionic operators we employ the Hubbard-Stratonovich transformation
\bq \label{HS}
&&\!\!\!\!\!\!\!\exp\left[\int d^{3}x \;
\frac{g^{2}}{2}\;j^{a\mu}j^a_{\mu}\right] =  \\ 
&&\int
\mathcal{D}a_{\mu}\exp\left[-\int d^{3}x \;
\mathrm{tr}\left(\frac{1}{2}a^{\mu}a_{\mu}+g\;
j^{\mu}a_{\mu}\right)\right],
\nonumber
\eq
which introduces the vector field $a_{\mu} = a^a_\mu T^a$. At this point we can integrate out the fermions that now appear quadratically. The resulting effective action is given by \cite{Seme,Red}
 \begin{eqnarray}
S_\mathrm{eff}[a]=&&\!\!\!\!\!-\ln \det \left(\displaystyle{\not}\partial-M+g\,\displaystyle{\not}a\right) \nonumber\\
=&&\!\!\!\!\!\frac{i}{8\pi} \frac{M}{|M|}\!\int \!\! d^{3}x\,L_\mathrm{CS}[a] + \!\mathcal{O}\left(\frac{\partial}{M}\right),
\label{eqn:eff}
\end{eqnarray}
where
 \begin{eqnarray}
L_\mathrm{CS}[a]=g^{2}\epsilon^{\lambda\mu\nu}\mathrm{tr} \!\left(a_{\lambda} \partial_\mu a_{\nu}+\frac{2}{3} \,g\,a_{\lambda} a_{\mu}a_{\nu}\right).
\label{eqn:cs}
\end{eqnarray}
The term $\mathcal{O}\left(\frac{\partial}{M}\right)$ has a negligible contribution to the low energy behaviour that we are interested in. Note that the action $S_\mathrm{eff}[a]$ is not gauge invariant for large gauge transformations \cite{Red}. It is possible to cure this global gauge anomaly by introducing a gauge-invariant regularisation such as the Pauli-Villars one \cite{Red,Ryu}. In this scheme the regularised action $S_\mathrm{eff}^{R}[a]=S_\mathrm{eff}[a]-\lim_{M_{0}^{2}\rightarrow \infty} S_\mathrm{eff}[a](M_{0})$ is given by
\begin{eqnarray}
S_\mathrm{eff}^{R}[a]=\lim_{M_{0}^{2}\rightarrow \infty}\frac{1}{2}\left(\frac{M}{|M|}-\frac{M_{0}}{|M_{0}|}\right) 
\frac{i}{4\pi}\int d^{3}x\;L_\mathrm{CS}[a].
\label{eqn:reneff}
\end{eqnarray}
When $\mathrm{sign}(M_{0})=-\mathrm{sign}(M)$, we obtain the standard non-Abelian Chern-Simons action with level $k=1$~\cite{Pisarski,Dunne}. It is worth noticing that changing the value of the coefficient $g$ in (\ref{eqn:cs}) does not change the value of the level of the non-Abelian theory~\cite{Pisarski,Dunne}. To simplify the next calculations we rescale $a_{\mu}\rightarrow a_{\mu}/g$ and take $M$ positive. 

Still the total action is not gauge invariant due to the $\int d^{3}x \;\mathrm{tr}(a^{\mu}a_{\mu})$ term in (\ref{HS}). It is possible to recast the total action in terms of a gauge invariant and renormalisable theory by introducing the interpolating action \cite{Deser,Karlhede}
\begin{eqnarray}
S_\mathrm{I}[a, A]=\int d^{3}x
\left\{\frac{1}{2 g^{2}} \mathrm{tr}\; a^{\mu}a_{\mu}+ \right. \nonumber \,\,\,\,\,\,\,\, \\ 
\left.
\frac{i}{2 \pi} \epsilon^{\mu\nu\lambda} \mathrm{tr}\;
a_{\mu}
\left[F_{\nu\lambda}(A)+\; A_{\nu}a_{\lambda}\right]+ \right. \nonumber \\ \left. \frac{i}{4 \pi}\,\epsilon^{\lambda\mu\nu}
\mathrm{tr} \left( A_{\lambda}
\partial_\mu A_{\nu}+\frac{2}{3}
A_{\lambda} A_{\mu}A_{\nu}\right) \right\}.
\label{eqn:inter}
\end{eqnarray}
If we shift the vector potential $A_{\mu}= A^a_\mu T^a$ as $A_{\mu}=\bar{A}_{\mu}-a_{\mu}$ and then integrate over $\bar{A}_{\mu}$, we find that the corresponding partition function becomes $Z_\mathrm{I} \approx  Z_\mathrm{Th}$, where the approximation is due to neglecting the $\mathcal{O}\left(\frac{\partial}{M}\right)$ term. If, on the other hand, we directly perform the $a_{\mu}$ integration in $Z_\mathrm{I}$ we obtain the following partition function
\begin{eqnarray}\label{FCS}
Z_\mathrm{FCS} \!\!\!\!\! &&=\!\! \int \!\mathcal{D} A_{\mu} \exp \left[\!-\! \int \!d^{3}x\,
\frac{g^{2}}{2 \pi^{2}}\, \mathrm{tr}\, ({}^*\! F_{\mu} S^{\mu\nu} {}^*\!F_{\nu})- \right.\nonumber\\ &&\left. -\frac{i}{4 \pi}\int \!d^{3}x\,
 \epsilon^{\lambda\mu\nu}
\mathrm{tr} \left( A_{\lambda}
\partial_\mu A_{\nu}+\frac{2}{3}
A_{\lambda} A_{\mu}A_{\nu}\right)\right],\,\,\,\,\,\,\,\,\,
\label{eqn:FCS}
\end{eqnarray}
where $S^{\mu\nu}=(\delta^{\mu\nu}+\frac{i g^{2}}{\pi}
\epsilon^{\mu\nu\lambda}A_{\lambda})^{-1}$ and ${}^*F_\mu = {1\over 2} \epsilon_{\mu \nu \lambda} F^{\nu \lambda}$. The first term of this action is a non-Abelian gauge theory that does not admit direct interpretation. The second term is the SU(2) Chern-Simons theory at level $k=1$ that gives mass to the gauge field and a finite correlation length $\xi$. As a result the large distance behaviour compared to $\xi$ is dominated by the Chern-Simons term with the contribution of the first term decaying exponentially fast away from the sources. 

\begin{figure}
\includegraphics[width=0.3\textwidth]{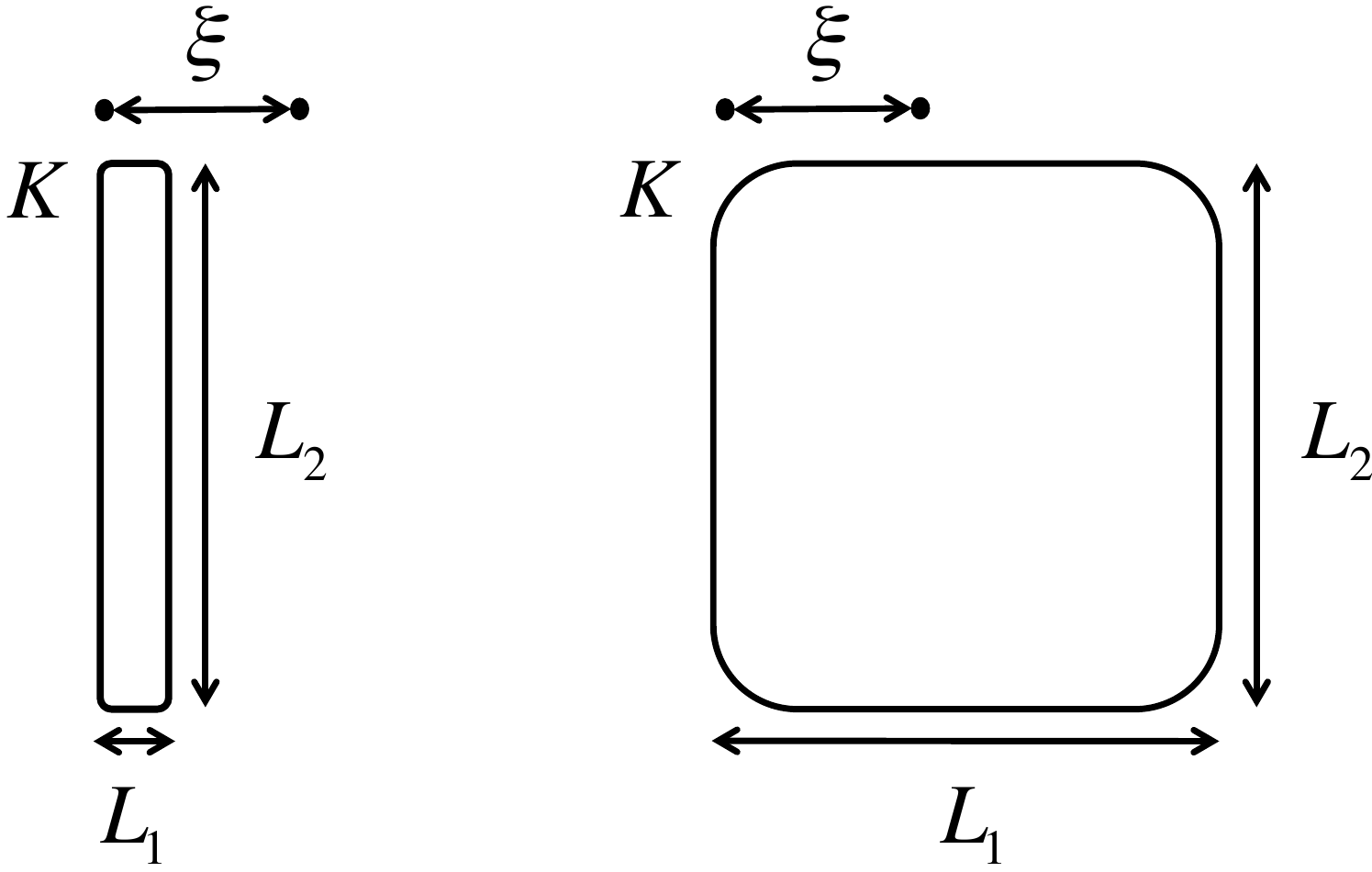}
\caption{\label{fig:Loops} Examples of a rectangular loops $K$ with area $A_K=L_1L_2$. (Left) When $L_1$ is small, of the order of the correlation length $\xi$, then the short distance behaviour of our model is given by (\ref{YM}). Then, confinement is manifested by the area-law of the Wilson loop observable, $\langle W(K)\rangle \approx e^{- \sigma A_{K}}$, with string tension $\sigma\sim {g^{-4}}$ \cite{Karabali}. (Right) When both $L_1$ and $L_2$ are large compared to the correlation length $\xi \sim g^{2}$ of the theory then the long distance behaviour is given by (\ref{CS}). Then $\langle W(K) \rangle = V_K(q)$, where $V_K(q)$ is the Jones polynomial of the loop $K$ with variable $q$. For the simple loop considered here it is $V_K(q)=1$. This is in stark contrast to the confining regime where $\langle W(K) \rangle$ tends to zero as $A_K$ increases.}
\end{figure}

The partition function (\ref{eqn:FCS}) describes our model for any value of $g$. Consider now the limit $g^{2}\ll 1$, where $S^{\mu\nu} \sim \delta^{\mu\nu}$~\cite{Fradkin2}. In this limit the short distance behaviour compared to $\xi$ of the FCS theory is described by the SU(2) Yang-Mills action 
\begin{eqnarray}\label{YM}
S_\mathrm{YM}[A]= \frac{g^{2}}{8 \pi^{2}}\! \int \!d^{3}x
\;\mathrm{tr}\,  F_{\mu\nu} F^{\mu\nu}
\end{eqnarray}
and (\ref{eqn:FCS}) defines a topologically massive gauge theory~\cite{Jackiw,Pisarski}.
Thus the original field theory, after the interpolating procedure, becomes the gauge invariant Yang-Mills-Chern-Simons theory in the limit $g^{2}\ll1$ and large mass $M$.
In particular the $(2+1)$-dimensional Yang-Mills theory supports confinement, one of the most intriguing challenges in high energy physics. Confinement can explain why free quarks cannot be experimentally detected. Nevertheless, this behaviour is analytically intractable to prove in $3+1$ dimensions \cite{Hooft}. To probe this property of (\ref{YM}) we introduce the Wilson loop operator. It is given by
\begin{eqnarray}\label{WilsonL}
W(K)=\mathrm{tr}\; \emph{P}\; e^{i\oint_{K}dx^{\mu}A_{\mu}},
\end{eqnarray}
where $\emph{P}$ denotes the path ordering necessary for non-Abelian theories, the trace is taken in the representation of the SU(2) algebra (taken here to be the fundamental) and $K$ is a given loop. Confinement is manifested by the area-law behaviour, 
\be
\langle W(K)\rangle \approx e^{- \sigma A_{K}}, 
\label{eqn:conf}
\ee
where $A_{K}$ is the area enclosed by the loop $K$ and the constant $\sigma$ is the string tension. An analytic expression for the string tension $\sigma$ has been derived for the $(2+1)$-dimensional SU(2) Yang-Mills theory, given by $\sigma\sim {g^{-4}}$ \cite{Karabali}. Loops $K$ that can probe this short-distance regime are shown in Fig. \ref{fig:Loops} (Left).
It is important to remark that there exists just a single quantum phase with different long- and short-range behaviours. Indeed
the behaviour of the Wilson loop changes drastically when we consider distances where the Chern-Simons term becomes relevant. This is achieved when we consider loops $K$ with geometric characteristics that are large compared to the correlation length $\xi$ of the system, as shown in Fig. \ref{fig:Loops} (Right). In this large-distance/small-energies regime we can ignore the Yang-Mills term and consider exclusively the Chern-Simons term that gives rise to a topological behaviour. This is the regime that we consider next.

We now extend Hamiltonian (\ref{Ham1}) in order to obtain theories with general integer $k$ that can support non-Abelian anyons. This is possible by introducing more than one species of spin-$1/2$ fermions. To illustrate that we start by considering $N$ copies of the model and parameterising the fermionic species by the index $\beta=1,...,N$. Then we modify the interaction term to obtain the SU(2) Thirring model with $N$ fermion species, namely
\be
H =\int d^3 x \Big[ \sum _{\beta=1}^N\psi^\dagger_{\beta} (\gamma_z\GAMMA\cdot \p+\gamma_z M)\psi_{\beta} + {g^2 \over 2} J^{a\mu} J^a_{\mu} \Big],
\label{eqn:extended}
\ee
where $\psi_{\beta} = (\psi_{\beta\uparrow},\psi_{\beta\downarrow})^T = (b_{\beta\uparrow},w_{\beta\uparrow},b_{\beta\downarrow},w_{\beta\downarrow})^T$, $ J^{a \mu} = \sum _{\beta=1}^N j^{a\mu}_\beta$ and $j^{a\mu}_\beta =\overline \psi_\beta T^a\gamma^\mu \psi_\beta$. This new interaction can be directly given in terms of the tight-binding fermions, $b_{\beta s}$ and $w_{\beta s}$, for $\beta =1,...,N$ and $s=\uparrow,\downarrow$. It represents a spin non-preserving interaction because it mixes the different fermionic species, as shown in Fig. \ref{fig:interactions} (Right). By performing the same bosonisation procedure as in the case of single species we obtain the effective action
\begin{eqnarray}
e^{-S_\mathrm{eff}[a]}\!\!&=&\!\!\!\!\int\!\mathcal{D}\overline{\psi}_{\beta}\mathcal{D}\psi_{\beta} \exp\left[ - \!\int d^{3}x
\; \overline{\psi}_{\beta}\left(\displaystyle{\not}\partial-M+
\displaystyle{\not}a\right)\psi_{\beta}\right]\nonumber \\
\!\!&=&\!\!\left[\det
\left(\displaystyle{\not}\partial-M+ \displaystyle{\not}a\right)\right]^{N},
\label{eqn:Nspecies}
\end{eqnarray}
where now the $N$-th power of the determinant arises. We can introduce an interpolating action, similar to (\ref{eqn:FCS}), to obtain the corresponding dual model. The small distance properties of this model, for $g^2\ll 1$, is described by the Yang-Mills term (\ref{YM}) multiplied now by $N^2$.  The large distance behaviour of (\ref{eqn:Nspecies}) is given by
\begin{eqnarray}\label{CS}
S_\mathrm{CS}[A]=\frac{N i}{4 \pi }\!\int \!d^{3}x\;
 \epsilon^{\lambda\mu\nu}
\mathrm{tr} \!\left( A_{\lambda}
\partial_\mu A_{\nu}+\frac{2}{3}
A_{\lambda} A_{\mu}A_{\nu}\right)\!,
\end{eqnarray}
which is the SU(2) level $k=N$ Chern-Simons theory. For $N=2$ it supports non-Abelian Ising anyons \cite{Wilczek}, which behave similarly to Majorana fermions. For $N=3$ Fibonacci anyons are supported, which are universal for quantum computation \cite{Freedman}. 

Witten showed that for the non-Abelian Chern-Simons theory the expectation value of the Wilson loop operator is given in terms of the Jones polynomial \cite{Witten}
\be
\langle W(K) \rangle = V_K(q)
\label{eqn:jones}
\ee
This relation holds for any link $K$ with possibly many strands. The Jones polynomial $V_K(q)$ has variable $q=\exp(2\pi i/(k+2))$, where $k$ is the level of Chern-Simons theory. It is a topological invariant of the link $K$, i.e. it does not depend on its geometrical characteristics, but only on its knottiness. For the case where $K$ is a single unknotted loop we have $V_K(q)=1$. In other words the expectation value of the corresponding Wilson loop is a constant, independent on the size of $K$. This statement is exact for large enough $K$ compared to the correlation length of the microscopic model, as shown in Fig. \ref{fig:Loops} (Right). This is in stark contrast to the short-distance behaviour of the model. The condition $\langle W(K) \rangle = 1$ for loops $K$ of any size and position is a witness of the model's topological order provided its ground state is not a trivial product state \cite{Giandomenico}.

It is worth noting that we do not have a direct way to measure the Wilson loop in terms of fermionic observables as we did in the Abelian case \cite{Giandomenico}. Nevertheless, it is possible to probe the topological order of the model through its behaviour at the boundary. For large characteristic geometries of the boundary so that short range correlations do not get involved the topological properties of the model can be isolated. It was shown in \cite{Witten, Elitzur} that the SU(2) level $k$ Chern-Simons bulk theory induces at its edge the SU(2) Wess-Zumino-Witten model, which is a conformal field theory \cite{WittenWZW}. To probe the Wess-Zumino-Witten model it is possible to measure the thermal currents at the boundary. For this model the thermal conductance $\emph{K}_{Q}$ of the edge modes is given by \cite{kane, cappe}
\begin{eqnarray}
\emph{K}_{Q}=\frac{\partial J_{Q}}{\partial T}=\frac{\pi}{6}\;c\; k_{B}^{2}T,
\label{therma}
\end{eqnarray}
which holds in the low temperature limit $T\rightarrow 0$. Here $J_{Q}$ is the thermal current carried by the edge modes, $k_{B}$ is the Boltzmann constant and $c=\frac{3k}{k+2}$ is the corresponding central charge. By employing (\ref{therma}) we can evaluate the level $k$ of the theory and thus determinate the particular species of anyons present in our model \cite{Pachos12}. The physical realisation of our model could be performed with cold atom methods proposed in \cite{Alba, Cirac2, Spielman, LewensteinBook, Brennen, Burrello, Das Sarma, Sato}, while a possible method to detect the chiral edge states is given in \cite{Goldman}.

{\em Acknowledgements:--} JKP would like to thank Stefanos Kourtis for inspiring conversations. This work was supported by EPSRC.


\begin{thebibliography}{25}



\bibitem{Gomes}
M. Gomes, R.S. Mendes, R.F. Ribeiro and A.J. da Silva, Phys. Rev. D {\bf 43}, 3516 (1991).

\bibitem{Fradkin}
E. Fradkin and F.A. Schaposnik, Phys. Lett. B {\bf 338}, 253 (1994).

\bibitem{Fradkin2}
N. Bralic, E. Fradkin, V. Manias and F.A. Schaposnik, Nucl. Phys. B {\bf 446}, 144 (1995).

\bibitem{Lewenstein}
L. Tagliacozzo, A. Celi, P. Orland, M.W. Mitchell and M. Lewenstein, Nature Communications {\bf 4}, 2615 (2013).

\bibitem{Zoller}
D. Banerjee, M. Bogli, M. Dalmonte, E. Rico, P. Stebler, U.-J. Wiese
and P. Zoller, Phys. Rev. Lett. {\bf 110}, 125303 (2013).

\bibitem{Cirac}
E. Zohar, J.I. Cirac and B. Reznik, Phys. Rev. Lett. {\bf 110}, 125304 (2013).

\bibitem{Pachos12}
J.K. Pachos, {\em Introduction to Topological Quantum Computation}, Cambridge University Press (2012).

\bibitem{Shapo2}
C.D. Fosco, G.L. Rossini and F.A. Schaposnik, Phys. Rev. D {\bf 56}, 6547 (1997).

\bibitem{Shapo3}
C.D. Fosco, G.L. Rossini and F.A. Schaposnik, Phys. Rev. D {\bf 59}, 085012 (1999).

\bibitem{Giandomenico}
G. Palumbo and J.K. Pachos, Phys. Rev. Lett. {\bf 110}, 211603 (2013).

\bibitem{Ryu}
A.P. Schnyder, S. Ryu, A. Furusaki and A.W.W. Ludwig, Phys. Rev. B {\bf 78}, 195125 (2008).

\bibitem{Parisi}
G. Parisi, Nucl. Phys. B {\bf 100}, 368 (1975).

\bibitem{Gies}
H. Gies and L. Janssen, Phys. Rev. D {\bf 82}, 085018 (2010).

\bibitem{Seme}
A.J. Niemi and G.W. Semenoff, Phys. Rev. Lett. {\bf 51}, 2077 (1983).

\bibitem{Red}
A.N. Redlich, Phys. Rev. D {\bf 29}, 2366 (1984).


\bibitem{Pisarski}
R.D. Pisarski and S. Rao, Phys. Rev. D {\bf 32}, 2081 (1985).

\bibitem{Dunne}
G.V. Dunne, Les Houches - Ecole d'Ete de Physique Theorique {\bf 69}, 177 (1999).


\bibitem{Deser}
S. Deser and R. Jackiw, Phys. Lett. B {\bf 139}, 371 (1984).

\bibitem{Karlhede}
A. Karlhede, U. Lindstrom, M. Rocek and P. Van Nieuwenhuizen, Phys. Lett. B {\bf 186}, 96 (1987).

\bibitem{Jackiw}
S. Deser, R. Jackiw and S. Templeton, Phys. Rev. Lett. {\bf 48}, 975 (1982).


\bibitem{Hooft}
G. 't Hooft, Nucl. Phys. B {\bf 138}, 1 (1978).



\bibitem{Karabali}
D. Karabali, C. Kim and V.P. Nair, Phys. Lett. B {\bf 434}, 103 (1998).



\bibitem{Wilczek}
E. Fradkin, C. Nayak, A. Tsvelik and F. Wilczek, Nucl. Phys. B {\bf 516}, 704 (1998).


\bibitem{Freedman}
M. Freedman, M. Larsen and Z. Wang, Comm. Math. Phys. {\bf 228}, 177 (2002).

\bibitem{Witten}
E. Witten, Comm. Math. Phys. {\bf 121}, 351 (1989).



\bibitem{Elitzur}
S. Elitzur, G. Moore, A. Schwimmer and N. Seiberg, Nucl. Phys. B {\bf 326}, 108 (1989).

\bibitem{WittenWZW}
E. Witten, Comm. Math. Phys. {\bf 92}, 455 (1984).

\bibitem{kane}
C.L. Kane and M.P.A. Fisher, Phys. Rev. B {\bf 55}, 15832 (1997).

\bibitem{cappe}
A. Cappelli, M. Huerta and G.R. Zemba, Nucl. Phys. B {\bf 636}, 568 (2002).





\bibitem{Alba}
E. Alba, X. Fernandez-Gonzalvo, J. Mur-Petit, J. K. Pachos and J. J. Garcia-Ripoll, Phys. Rev. Lett. {\bf 107}, 235301 (2011).

\bibitem{Cirac2}
J. I. Cirac, P. Maraner and J. K. Pachos, Phys. Rev. Lett. {\bf 105}, 190403 (2010).

\bibitem{Brennen}
M. Aguado, G.K. Brennen, F. Verstraete and J.I. Cirac, Phys. Rev. Lett. {\bf 101}, 260501 (2008).

\bibitem{Burrello}
M. Burrello and A. Trombettoni, Phys. Rev. Lett. {\bf 105}, 125304 (2010).

\bibitem{Das Sarma}
S. Tewari, S. Das Sarma, C. Nayak, C. Zhang and P. Zoller, Phys. Rev. Lett. {\bf 98}, 010506 (2007).

\bibitem{Sato}
M. Sato, Y. Takahashi, and S. Fujimoto, Phys. Rev. Lett. {\bf 103}, 020401 (2009).

\bibitem{Spielman}
N. Goldman, E. Anisimovas, F. Gerbier, P. Ohberg, I.B. Spielman and G. Juzeliunas, New J. Phys. {\bf 15}, 13025 (2013).

\bibitem{LewensteinBook}
M. Lewenstein, A. Sanpera and V. Ahufinger, {\em Ultracold Atoms in Optical Lattices: Simulating quantum many-body systems}, Oxford University Press (2012).

\bibitem{Goldman}
N. Goldman, J. Dalibard, A. Dauphin, F. Gerbier, M. Lewenstein, P. Zoller and I.B. Spielman, PNAS {\bf 110}, 6736 (2013).

\end{thebibliography}
\end{document}